\documentclass[aps,prb,reprint,showpacs]{revtex4-1}
\usepackage{amsmath}
\usepackage{amssymb}
\usepackage{amsbsy}
\usepackage{graphicx}
\usepackage{multirow}
\usepackage{undertilde}
\usepackage{natbib}
\usepackage[bookmarks,colorlinks=true,linkcolor=blue,citecolor=blue,urlcolor=blue]{hyperref}

\newcommand{\ud}{\text{d}}
\newcommand{\iu}{\text{i}}
\newcommand{\Tr}{\text{Tr}\,}
\newcommand{\ImTr}{\text{Im}\,\text{Tr}}
\newcommand{\inv}[1]{^{\text{--}#1}}
\newcommand{\TENSOR}[3]{\boldsymbol{#1}_{#2}^{#3}}

\begin{document}
\title{Anisotropic spin--spin correlations in Mn$_1$/X(111), with X = Pd, Pt, Ag and Au}
\date{\today}

\author{M \surname{dos Santos Dias}}
\email{m.d-s.dias@warwick.ac.uk}
\author{JB Staunton}
\affiliation{Department of Physics, University of Warwick, Coventry CV4 7AL, United Kingdom}
\author{A Deak}
\author{L Szunyogh}
\affiliation{Department of Theoretical Physics, Budapest University of Technology and Economics, Budapest, Hungary}

\begin{abstract}
We present a finite--temperature theory of the anisotropic spin--spin correlations in magnetic metallic monolayers, deposited on a suitable substrate. The `spins' are the local moments set up by the itinerant electrons, and the key concept is the relativistic disordered local moment state, which represents the paramagnetic state of a set of local moments. The spin--spin correlations between these local moments are then extracted using the linear response formalism. The anisotropy is included in a fully relativistic treatment, based on the Dirac equation, and has a qualitative impact on noncollinear magnetic states, by lifting their chiral degeneracy. The theory is applied to Mn monolayers on the hexagonal (111) surfaces of Pd, Pt, Ag and Au. The presence of competing exchange interactions is highlighted by choosing different substrates, which favour either the row--wise antiferromagnetic state or the chiral triangular N\'eel state. We correlate the electronic structure with the magnetic properties, by comparing filled with partially filled substrate d--bands, and low vs high atomic number. The disagreement between theory and experiment for Mn$_1$/Ag(111) is addressed, and the nature of the magnetic domains found experimentally is suggested to be chiral.
\end{abstract}

\maketitle

\section{Introduction}
A thorough theoretical description of the magnetic properties of nanostructures remains to be achieved, despite the intense effort dedicated to this task. This is of intrinsic interest not only for fundamental theoretical and experimental physics\cite{Wiesendanger_review,Kirschner_review}, but also for the practical advances that make use of these discoveries to improve our current technology. In this paper we wish to present a contribution to this goal, by outlining our finite--temperature theory of the anisotropic spin--spin correlations in magnetic metallic monolayers (MLs), deposited on a suitable substrate, and demonstrating its application to Mn MLs on the hexagonal fcc (111) surfaces of Pd, Pt, Ag and Au. In these systems competing interactions delicately balance to favour either row--wise antiferromagnetic (AF) or triangular AF groundstates. Very complex unidirectional anisotropies, of Dzyaloshinskii--Moriya (DM) type, then conspire to lift the chiral degeneracy of the noncollinear state. We extract all this information from the analysis of the high--temperature paramagnetic susceptibility, which has the spin--spin correlation function as its main ingredient.

In itinerant systems there is a mutual feedback between the electronic and the magnetic properties. The nature of the magnetic state affects the electronic motions, which in turn determine the magnetic properties. At T = 0 K, first--principles theories such as Density Functional Theory (DFT) provide a quantitative account of the groundstate properties\cite{Kubler_book}. At finite temperature, however, the link between electronic structure and magnetism is usually broken, by mapping the T = 0 K information to some effective magnetic model, which properties are then computed either by analytical or numerical techniques\cite{Pajda_Jijs}.

The theory we present in this paper restores this link, by making use of the Disordered Local Moment (DLM) picture\cite{Gyorffy_DLM,Staunton_Onsager,Staunton_TempAnis}. In the DLM state the relevant degrees of freedom are the local spin quantisation axes associated with each lattice site, which can have associated local spin moments under favourable conditions. Their orientations are assigned according to a specified probability distribution, and the electrons move through this lattice of disordered moments in a mean--field fashion, in keeping with the spirit of DFT. The formulation is completely general, and can be applied either to bulk or thin films, for which the multi--sublattice or multilayer descriptions apply\cite{Staunton_susc,Razee_S1more,Staunton_TempAnis,Burusz_S1}.

We focus on the high--temperature paramagnetic (PM) state, for which a classical description of the local moments is expected to be adequate. Using linear response theory, we derive the pair correlation function between local moments, which yields spin--spin correlations similar to those obtained in the Random Phase Approximation\cite{Gyorffy_DLM} (RPA). Previous work along these lines focused on the interlayer couplings\cite{Razee_S1more}, and we expand on it by also treating intralayer couplings, both in real and reciprocal space. The fundamental extension of the theory presented here is the relativistic formulation, going beyond the isotropic correlations by naturally incorporating anisotropic effects, which determine the real--space magnetic structure and lift degeneracies, such as chirality. In low--dimensional systems it is the magnetic anisotropy which stabilises the ordered state against spin fluctuations, and so it is a crucial ingredient of any theory aimed at describing nanostructures.

The theory is then applied to Mn MLs on the hexagonal (111) surfaces of Pd, Pt, Ag and Au. Our motivation stems from recent experiments performed for Mn$_1$/Ag(111), which were interpreted as evidence of a triangular AF state\cite{Gao_MnAg}, in contrast to the predictions from previous electronic structure calculations, which pointed to a row--by--row AF state\cite{Heinze_MnAg,Kruger_MnAg}, and even ruled out a more exotic 3D structure\cite{Heinze_MnAg}, such as the one predicted for Mn$_1$/Cu(111)\cite{Kurz_MnCu111}. By investigating the properties of Mn MLs on these four closely related substrates, we uncovered a consistent picture of competing magnetic interactions leading to frustration and different magnetic groundstates very close in energy, coupled to very complex DM--like anisotropies, which favour chiral triangular AF states. Our findings are consistent with other published calculations, and suggest that further theoretical and experimental work is required to shed light on these deceptively simple systems.

We begin by presenting our new theory of the anisotropic spin--spin correlations in magnetic metallic MLs. The following section contains the details of our numerical calculations, which are presented afterwards. We finish by summarising our results and presenting an outlook on further work.

\section{Theoretical foundation}
\subsection{Local moments in metals}
Local magnetic moments in metals are an emergent property\cite{Moriya_book,Gyorffy_DLM,Staunton_Onsager}. Due to electron--electron interactions, an electron in an itinerant system will have an energetic preference to move between lattice sites which have an overall spin polarisation more or less aligned with its own spin. If the orientations of the local spin polarisation on the lattice sites vary more slowly than the timescale of the electronic motions, collective degrees of freedom can be meaningfully assigned to those lattice sites, which we label `local moments'. The magnitude of these local moments will vary in time as fast as the electrons hopping to and from each site, but on the timescale of the precessional motions a well--defined time--averaged magnitude is apparent.

Good local moment systems are those for which the magnitude of the magnetic moments is fairly insensitive to the local and global magnetic state\cite{Mryasov_model,Sandratskii_model}. For these systems one can make use of the Rigid Spin Approximation (RSA). In DFT language it means that the exchange--correlation magnetic field entering the Kohn--Sham equations is the same for any orientation of the local moment. It is an approximation because there will always be feedback between the electronic and the magnetic properties in an itinerant system, however it has been shown in the past that for many systems it holds very well, for example by performing spin--spiral calculations\cite{Heinze_MnAg}. We shall make use of the RSA in our construction of the DLM state. It should be clear, however, that the electronic structure in the RSA is still different for different magnetic states, and even more so for the DLM state, due to the ensemble averages over the local moment orientations.

\subsection{Statistical mechanics of localised moments}
Let us now develop the meaning of our theory of the anisotropic spin--spin correlation function by considering the statistical mechanics of a local moment system. The Hamiltonian for the system is written as
\begin{equation}
  \mathcal{H} = \mathcal{H}_{\text{int}} + \mathcal{H}_{\text{ext}} =  \mathcal{H}_{\text{int}}(\{\hat{e}\}) - \sum_{i=1}^N \hat{e}_i\cdot\vec{H}_i
\end{equation}
The relevant degrees of freedom are the orientations of the local spin quantisation axes, written as unit vectors $\hat{e}_i$. The first term is the interaction Hamiltonian, describing the coupling between the local moments, and is left unspecified. The second term is the Zeeman energy due to coupling to the external applied fields $\vec{H}_i$.

The free energy is given by
\begin{equation}
  e^{-\beta\mathcal{F}} = \mathcal{Z} = \prod_{k=1}^N\!\int\!\!\ud\hat{e}_k\,e^{-\beta\mathcal{H}}
\equiv \int\!\!\ud\{\hat{e}\}\,e^{-\beta\mathcal{H}}
\end{equation}
which also defines the partition function, $\mathcal{Z}$. Here $1/\beta = k_{\text{B}} T$, with $k_{\text{B}}$ Boltzmann's constant and $T$ the temperature. Then the local order parameter, $\vec{m}_i \equiv \langle\,\hat{e}_i\,\rangle$, is obtained from
\begin{equation}
  \vec{m}_i = -\frac{\partial\mathcal{F}}{\partial\vec{H}_i} =  \int\!\!\ud\{\hat{e}\}\,\frac{e^{-\beta\mathcal{H}}}{\mathcal{Z}}\,\hat{e}_i = \int\!\!\ud\{\hat{e}\}\,P(\{\hat{e}\})\,\hat{e}_i
\end{equation}
which also introduces the probability distribution, $P(\{\hat{e}\})$, and the magnitude of the local moments is omitted. It can be reintroduced in the final expression for the susceptibility, if desired.

The magnetic susceptibility then follows
\begin{equation}
  \TENSOR{\chi}{ij}{×} = \frac{\partial\vec{m}_i}{\partial\vec{H}_j} = \beta\left(\langle\,\hat{e}_i\,\hat{e}_j\,\rangle - \langle\,\hat{e}_i\,\rangle\langle\,\hat{e}_j\,\rangle\right) \vspace{-0.5em}
\end{equation}
The boldface signifies a $3\times3$ tensor in Cartesian components. This expression defines a spin--spin pair correlation function, and connects it to the magnetic susceptibility --- it is a special case of the fluctuation--dissipation theorem\cite{White_book}.

To proceed with the analysis some approximations are required. It is already apparent that for the two quantities of interest, the local order parameter and the magnetic susceptibility, full knowledge of the complete probability distribution is not necessary, and that only reduced probability distributions are required:
\begin{align}
  \langle\,\hat{e}_i\,\rangle &= \int\!\!\ud\hat{e}_i\,P_1(\hat{e}_i)\,\hat{e}_i \\
  \langle\,\hat{e}_i\,\hat{e}_j\,\rangle &= \int\!\!\ud\hat{e}_i\!\int\!\!\ud\hat{e}_j\,P_2(\hat{e}_i,\hat{e}_j)\,\hat{e}_i\,\hat{e}_j
\end{align}
where, for instance, $P_2(\hat{e}_i,\hat{e}_j) = \prod_{k \neq i,j}\!\int\!\ud\hat{e}_k\,P(\{\hat{e}\})$. Thus a useful approximation may focus on obtaining these restricted averages.

The other obstacle is the need to handle an arbitrary, possibly highly complicated, interaction Hamiltonian, furnished by the first--principles electronic structure. A very successful scheme to address this matter is that of variational statistical mechanics, in which a trial Hamiltonian, $\mathcal{H}_0(\{\hat{e}\})$, is chosen and its parameters determined from a variational bound on the free energy. This can be obtained from the Feynman--Peierls--Bogoliubov inequality\cite{Feynman_inequality},
\begin{equation}
  \mathcal{F} \leq \mathcal{F}_0 + \langle\,\mathcal{H} - \mathcal{H}_0\,\rangle\;,\;\;\langle\,X\,\rangle = \int\!\!\ud\{\hat{e}\}\,\frac{e^{-\beta\mathcal{H}_0}}{\mathcal{Z}_0}\,X
\end{equation}
so that the ensemble averages are calculated using the probability distribution generated by the trial Hamiltonian. The variational parameters are then determined by ensuring the equality of the respective restricted averages\cite{Gyorffy_DLM},
\begin{align}
  \langle\,\mathcal{H}\,\rangle_{\hat{e}_i} - \langle\,\mathcal{H}\,\rangle &= \langle\,\mathcal{H}_0\,\rangle_{\hat{e}_i} - \langle\,\mathcal{H}_0\,\rangle \label{partavs}\\
  \langle\,\mathcal{H}\,\rangle_{\hat{e}_i,\hat{e}_j} - \langle\,\mathcal{H}\,\rangle &= \langle\,\mathcal{H}_0\,\rangle_{\hat{e}_i,\hat{e}_j} - \langle\,\mathcal{H}_0\,\rangle \\
&\hspace{0.5em}\vdots \nonumber
\end{align}
where all orientations are averaged over except those singled out, i.e., the orientations on sites $i$ and $j$ are kept fixed where stated.

For simplicity, we choose a site--diagonal trial Hamiltonian, $\mathcal{H}_0 = - \sum_i \vec{h}_i\cdot\hat{e}_i$. Then the so--called Weiss fields are given by
\begin{equation}
  \vec{h}_i = \vec{H}_i - \frac{3}{4\pi}\int\!\ud\hat{e}_i\,\hat{e}_i\,\langle\,\mathcal{H}\,\rangle_{\hat{e}_i}\;\text{ or }\; \vec{h}_i = \vec{H}_i - \frac{\partial\langle\,\mathcal{H}\,\rangle}{\partial\vec{m}_i}
\end{equation}
The first definition follows directly from Eq. (\ref{partavs}), and is computationally tractable. The second one arises from the direct minimisation of the trial free energy, and is conceptually more useful.

The local order parameter is then given by the Langevin function,
\begin{equation}
  \vec{m}_i = \left[\coth(\beta h_i) - \frac{1}{\beta h_i}\right]\!\hat{h}_i = L(\beta h_i)\,\hat{h}_i\;,\;\;\hat{h}_i = \frac{\vec{h}_i}{h_i}
\end{equation}
By taking its derivative with respect to external applied fields $\{\vec{H}\}$, and making use of
\begin{equation}
  \frac{\partial\vec{m}_i}{\partial\vec{H}_j} = \frac{\partial\vec{m}_i}{\partial\vec{h}_i}\frac{\partial\vec{h}_i}{\partial\vec{H}_j} \quad\text{and}\quad \frac{\partial h_i}{\partial\vec{h}_i} = \hat{h}_i
\end{equation}
the magnetic susceptibility quickly follows:
\begin{align}
  \TENSOR{\chi}{ij}{×} \equiv \frac{\partial\vec{m}_i}{\partial\vec{H}_j} = \TENSOR{\chi}{0,i}{×}\,\delta_{ij} + \TENSOR{\chi}{0,i}{×}\cdot\sum_{k\neq i}\TENSOR{\mathcal{S}}{ik}{}\cdot\TENSOR{\chi}{kj}{}
\end{align}
where we introduce our spin--spin direct correlation function\cite{Gyorffy_DLM,Staunton_susc},
\begin{equation}
  \TENSOR{\mathcal{S}}{ij}{\alpha\beta} = \frac{\partial\vec{h}_i^\alpha}{\partial\vec{m}_j^\beta} = -\frac{\partial^2\langle\,\mathcal{H}\,\rangle}{\partial\vec{m}_i^\alpha\partial\vec{m}_j^\beta}\;,\quad\alpha,\beta=x,y,z \label{s2def}
\end{equation}
which measures the interactions between the local moments. This is the key quantity for our first--principles calculations. The statistical mechanics approach we just outlined is completely general, and has been applied in other order--disorder contexts\cite{Gyorffy_ConcWaves,Staunton_SRO,Razee_MCAnSRO}.

The local susceptibility tensor is given by
\begin{align}
  \TENSOR{\chi}{0,i}{\alpha\beta} &= \beta \left[\left(L'(\beta h_i) - \frac{L(\beta h_i)}{\beta h_i}\right)\hat{h}_i^\alpha\hat{h}_i^\beta + \frac{L(\beta h_i)}{\beta h_i}\,\delta_{\alpha\beta}\right] \nonumber\\
&\longrightarrow \frac{\beta}{3}\,\delta_{\alpha\beta}\quad\text{(PM state)}
\end{align}
and can be recognised as the Langevin susceptibility of non--interacting local moments.

Performing a Legendre transform on the free energy, replacing the external applied fields by the local order parameters as independent variables, we obtain in the vicinity of a transition from the PM state into some ordered state
\begin{align}
  \mathcal{F}(\{\vec{m}\};T\!\rightarrow\!T_c) \approx \frac{1}{2}\sum_{i,j}\delta\vec{m}_i\cdot\TENSOR{\chi}{ij}{\text{--}1}\cdot\delta\vec{m}_j &+ \ldots \nonumber\\
= \frac{1}{2}\sum_{i,j}\delta\vec{m}_i\cdot\Big(3 k_{\text{B}}T\,\delta_{ij}\,\TENSOR{\mathcal{I}}{×}{×} - \TENSOR{\mathcal{S}}{ij}{}\Big)\cdot\delta\vec{m}_j &+ \ldots \label{free}
\end{align}

At sufficiently high temperature this quadratic form is positive definite, and so the minimum of the free energy is attained for $\{\delta\vec{m}\} = \{\vec{0}\}$, the PM state. The highest temperature at which this quadratic form is no longer positive definite will then be the transition temperature $T_c$ into a magnetically ordered state. It is the presence of interactions between the local moments, encoded in $\mathcal{S}_{ij}$, which drives this transition.

We proceed by diagonalising this quadratic form in a Fourier basis, using $\delta\vec{m}_i = \Omega_{\text{BZ}}\inv{1}\!\int\!\ud\vec{q}\,\exp(-\iu\vec{q}\cdot\vec{R}_i)\,\delta\vec{m}(\vec{q})\,$:
\begin{equation}
  \sum_{i,j}\delta\vec{m}_i\cdot\TENSOR{\mathcal{S}}{ij}{}\!\cdot\delta\vec{m}_j = \int\!\!\frac{\ud\vec{q}}{\Omega_{\text{BZ}}}\,\delta\vec{m}(-\vec{q})\cdot\TENSOR{\mathcal{S}}{}{}(\vec{q})\cdot\delta\vec{m}(\vec{q})
\end{equation}
where $\Omega_{\text{BZ}}$ is the Brillouin zone (BZ) volume/area, solving the $3\times3$ Hermitian eigenvalue problem for each $\vec{q}$:
\begin{equation}
  \TENSOR{\mathcal{S}}{}{}(\vec{q})\cdot\vec{u}_p(\vec{q}) = \sigma_p(\vec{q})\,\vec{u}_p(\vec{q})\;,\;\;p=1,2,3 \label{eigen}
\end{equation}
with eigenvalues $\sigma_p(\vec{q})$ and eigenvectors $\vec{u}_p(\vec{q})$, which can be chosen orthonormal, and define the basis in which $\TENSOR{\mathcal{S}}{}{}(\vec{q})$ is diagonal. Then $T_c$ is given by
\begin{equation}
   3 k_{\text{B}} T_{c} = \max \sigma_p(\vec{q}) \;\Longrightarrow\; \delta\vec{m}(\vec{q}) \propto \vec{u}_p(\vec{q}) \label{tc}
\end{equation}
and the associated eigenvector determines the magnetisation profile associated with the instability. All symmetry--related $\vec{q}$ vectors will have the same eigenvalues, and so in this single--$\vec{q}$ picture they are energetically degenerate.

The standard decomposition of a $3\times3$ Cartesian tensor will be used, and will be found to provide a natural interpretation of the information contained in $\TENSOR{\mathcal{S}}{ij}{}$. It can be written as
\begin{align}
  \TENSOR{\mathcal{S}}{ij}{} &= \TENSOR{\mathcal{S}}{ij}{\text{I}} + \TENSOR{\mathcal{S}}{ij}{\text{S}} + \TENSOR{\mathcal{S}}{ij}{\text{A}}\;,\;\;\TENSOR{\mathcal{S}}{ij}{\text{I}} = \frac{1}{3}\,\Tr \TENSOR{\mathcal{S}}{ij}{} \nonumber\\
\TENSOR{\mathcal{S}}{ij}{\text{S}} &= \frac{1}{2}\Big(\TENSOR{\mathcal{S}}{ij}{} + \TENSOR{\mathcal{S}}{ij}{T}\Big) - \TENSOR{\mathcal{S}}{ij}{\text{I}}\;,\;\;
\TENSOR{\mathcal{S}}{ij}{\text{A}} = \frac{1}{2}\Big(\TENSOR{\mathcal{S}}{ij}{} - \TENSOR{\mathcal{S}}{ij}{T}\Big) \label{decomp}
\end{align}

The isotropic part is the dominant contribution. From $\TENSOR{\mathcal{S}}{}{\text{I}}(\vec{q}) = \sum_j \cos(\vec{q}\cdot\vec{R}_{ij})\,\TENSOR{\mathcal{S}}{ij}{\text{I}}$ it can be seen that it will determine the modulation vector $\vec{q}$, with corrections from the anisotropic terms, if they are strong enough.

The antisymmetric part of the tensor can be cast into the form of an unidirectional, or DM--type, anisotropy:
\begin{equation}
  \sum_{i,j}\delta\vec{m}_i\cdot\TENSOR{\mathcal{S}}{ij}{\text{A}}\cdot\delta\vec{m}_j = \sum_{i,j}\vec{D}_{ij}\cdot(\delta\vec{m}_i\times\delta\vec{m}_j) \label{dmdef}
\end{equation}
From $\TENSOR{\mathcal{S}}{}{\text{A}}(\vec{q}) = \iu\sum_j \sin(\vec{q}\cdot\vec{R}_{ij})\,\TENSOR{\mathcal{S}}{ij}{\text{A}}$ it is apparent that in general this term will give rise to complex eigenvectors, and to the notion of chirality. This is demonstrated with a simple example in the next section, and with calculations for real systems later.

The remaining uniaxial anisotropy, $\TENSOR{\mathcal{S}}{ij}{\text{S}}$, determines the real space orientation of the local moments, being of pseudo--dipolar form. The familiar magnetostatic dipole--dipole interaction is of this form, and if it is added to the free energy it will be incorporated in this term.

To put flesh on these statistical mechanics bones we will now illustrate this approach for a generalised anisotropic Heisenberg model, and compare with more familiar treatments.

\subsection{Spin--spin correlations in the Heisenberg model}
Suppose that the hitherto unspecified interaction Hamiltonian is of the form of a generalised anisotropic Heisenberg model,
\begin{equation}
  \mathcal{H}_{\text{int}} = -\frac{1}{2}\sum_{i,j}\hat{e}_i\cdot\TENSOR{\mathcal{J}}{ij}{}\cdot\hat{e}_j = -\frac{1}{2}\sum_{i,j}\sum_{\alpha,\beta}\hat{e}_i^\alpha\,\TENSOR{\mathcal{J}}{ij}{\alpha\beta}\,\hat{e}_j^\beta
\end{equation}
with $\alpha,\beta=x,y,z$. Choosing the same trial Hamiltonian, we obtain for the Weiss fields
\begin{equation}
  \vec{h}_i = \vec{H}_i + \sum_{j \neq i}\TENSOR{\mathcal{J}}{ij}{}\cdot\vec{m}_j
\end{equation}
which reproduces the well--known mean field result. This particular choice of trial Hamiltonian will only capture the bilinear part of a generalised interaction Hamiltonian, as can be seen from its definition. In this model we can identify $\TENSOR{\mathcal{S}}{ij}{} \leftrightarrow \TENSOR{\mathcal{J}}{ij}{}$, and make immediate use of the discussion from the previous section.

The magnetic susceptibility tensor is then
\begin{equation}
  \TENSOR{\chi}{ij}{} = \TENSOR{\chi}{0,i}{}\,\delta_{ij} + \sum_{k \neq i}\TENSOR{\chi}{0,i}{}\cdot\TENSOR{\mathcal{J}}{ik}{}\cdot\TENSOR{\chi}{kj}{}
\end{equation}
Taking the lattice Fourier transform and inverting to solve for the susceptibility,
\begin{equation}
  \TENSOR{\chi}{}{}(\vec{q}) = \Big[(\TENSOR{\chi}{0}{})\inv{1} - \TENSOR{\mathcal{J}}{}{}(\vec{q})\,\Big]\inv{1} \!\!\xrightarrow[\text{PM}]{} \Big[3 k_{\text{B}} T \,\TENSOR{\mathcal{I}}{×}{×}- \TENSOR{\mathcal{J}}{}{}(\vec{q})\,\Big]\inv{1}
\end{equation}
and the paramagnetic limit is shown. Thus the mean field theory in an external applied field yields non--local correlations for the Heisenberg model, which are governed by the Heisenberg exchange interactions.

To finish this section we show an example of what the Fourier transformed Cartesian tensor might look like, for a particular modulation vector $\vec{q}$. Using Eq. (\ref{tc}) to interpret the tensor, along with $\vec{u}_i = \text{Re}[\,\exp(-\iu\vec{q}\cdot\vec{R}_i)\,\vec{u}(\vec{q})\,]$,
\begin{align}
&\TENSOR{\mathcal{J}}{}{}(\vec{q})\cdot\vec{u}_p(\vec{q}) = \sigma_p(\vec{q})\,\vec{u}_p(\vec{q})\;,\quad \TENSOR{\mathcal{J}}{}{}(\vec{q}) = 
\begin{pmatrix}
J_{\parallel} & \iu D & 0 \\
-\iu D & J_{\parallel} & 0 \\
0 & 0 & J_{\perp}
\end{pmatrix} \nonumber\\
&\Rightarrow \sigma_1(\vec{q}) = J_{\parallel} + D\;,\;\; \sigma_2(\vec{q}) = J_{\parallel} - D \;,\;\; \sigma_3(\vec{q}) = J_{\perp} \nonumber\\
&\rightarrow \vec{u}_1(\vec{q}) = \hat{x} - \iu\hat{y}\;,\;\;\vec{u}_i = \cos(\vec{q}\cdot\vec{R}_i)\,\hat{x} - \sin(\vec{q}\cdot\vec{R}_i)\,\hat{y} \nonumber\\
&\rightarrow \vec{u}_2(\vec{q}) = \hat{x} + \iu\hat{y}\;,\;\;\vec{u}_i = \cos(\vec{q}\cdot\vec{R}_i)\,\hat{x} + \sin(\vec{q}\cdot\vec{R}_i)\,\hat{y} \nonumber\\
&\rightarrow \vec{u}_3(\vec{q}) = \hat{z}\hspace{2.1em}\;,\;\;\vec{u}_i = \cos(\vec{q}\cdot\vec{R}_i)\,\hat{z} \label{example}
\end{align}
This depicts the case of a uniaxial exchange with a DM--type contribution equivalent to a DM vector along $z$. The two in--plane eigenvalues have complex eigenvectors, representing two possible chiralities, and their degeneracy is removed by the DM anisotropy. These chiral states are illustrated in the two bottom panels of Fig. \ref{magq}.

We now proceed to derive the magnetic interactions from first--principles electronic structure theory. First we must construct a description of the high temperature paramagnetic state, using the DLM state. Then, by carrying out the same type of analysis, the spin--spin correlations in the paramagnetic DLM state are derived. These take the place of the Heisenberg exchange parameters, but can be more general, as will be seen.

\subsection{The Disordered Local Moment state}
Our theory of the DLM state is based on DFT, in the Kohn--Sham picture, which is a single--particle mean--field theory. Each electron moves in the self--consistent field (SCF) generated by all other electrons. A local spin quantisation axis can then be naturally associated with each lattice site, according to the average orientation of the spin--only magnetic Kohn--Sham potential. The electronic properties are obtained from Multiple Scattering Theory\cite{Papanikolaou_KKR}, and the detailed derivation of the DLM equations has been presented before, for the relativistic\cite{Staunton_TempAnis} and non--relativistic\cite{Gyorffy_DLM} cases. In what follows only the essential definitions and equations will be collected, and we refer to the appropriate references for more information.

The Kohn--Sham potentials for a lattice site $i$ with the local spin quantisation axis oriented along $\hat{e}_i$ define the single--site t--matrix, $\underline{t}_i(\hat{e}_i;\varepsilon)$, for given energy $\varepsilon$, through the Lippmann--Schwinger equation. Making use of the RSA for spherical potentials, the t--matrix for an arbitrary orientation $\hat{e}_i$ is simply related to that of a given reference one, for example along the z--axis, by an appropriate rotation\cite{Staunton_TempAnis}:
\begin{equation}
  \underline{t}_i(\hat{e}_i;\varepsilon) = \underline{R}(\hat{e}_i)\,\underline{t}_i(\hat{z};\varepsilon)\,\underline{R}^\dagger(\hat{e}_i)
\end{equation}
Without SOC the spin components factorise, and the $t$--matrix takes on the form\cite{Gyorffy_DLM}
\begin{equation}
  \underline{t}_i(\hat{e}_i;\varepsilon) = \underline{t}_i^+(\varepsilon)\,\mathcal{I} + \underline{t}_i^-(\varepsilon)\,\vec{\sigma}\cdot\hat{e}_i
\end{equation}
with $\vec{\sigma}$ the vector of Pauli spin 1/2 matrices and $\mathcal{I}$ the identity matrix. It can be shown that then the only possible couplings between the orientations of the local moments at two different sites is some power of $\hat{e}_i\cdot\hat{e}_j$, which is isotropic in real space. Only by incorporating relativistic effects can this limitation be overcome\cite{Staunton_imp2}, and this is the essential new feature of our extension of the theory of spin--spin correlations in local moment systems.

Relativistic effects arise by solving the single--site problem using the Dirac instead of the Schr\"odinger equation\cite{Strange_Dirac,Strange_RelKKR}. The spin--only magnetic Kohn--Sham potential, which gives rise to the local spin moment, will then also generate a local orbital moment, through the SOC effect. The orbital moment is an induced property, resulting from the relativistic description, and is not an independent order parameter in our theory.

Propagation between lattice sites is described by making use of a suitably chosen reference system, for which the Green's function is known, $\underline{G}^0_{ij}(\varepsilon)$. The scattering path operator is then given by\cite{Gyorffy_SPO}
\begin{equation}
  \underline{\tau}_{ij}(\varepsilon) = \underline{t}_i(\varepsilon) + \underline{t}_i(\varepsilon)\sum_{k \neq i} \underline{G}^0_{ik}(\varepsilon)\,\underline{\tau}_{kj}(\varepsilon)
\end{equation}
and the integrated density of states (DOS) by Lloyd's formula\cite{Lloyd_formula},
\begin{equation}
  N(\varepsilon) = \frac{1}{\pi}\,\ImTr\,\log \underline{\underline{\tau}}(\varepsilon) \label{lloyd}
\end{equation}
where the double underlines stand for matrices in site and angular momentum labels, and the local moment orientations were omitted.

The resulting Hamiltonian, $\mathcal{H}(\{\hat{e}\})$, for an arbitrary set of local moment orientations, $\{\hat{e}\}$, is intractable. The theory progresses by invoking a simpler reference Hamiltonian, $\mathcal{H}_0 = -\sum_i \vec{h}_i\cdot\hat{e}_i$, and making use of the Feynman--Peierls--Bogoliubov inequality to determine the best Weiss fields, as described in the previous sections.

To carry out the restricted averages the Coherent Potential Approximation\cite{Soven_CPA} (CPA) is invoked. In this scheme we replace the true disordered medium by an effective translationally invariant one, such that the average properties of an electron are reproduced. In the single--site approximation one requires the effective single--site t--matrix, $\underline{\tilde t}_i(\varepsilon)$, which is chosen to satisfy\cite{Staunton_TempAnis}
\begin{equation}
  \int\!\ud\hat{e}_i\,P_i(\hat{e}_i)\,\underline{X}_i(\hat{e}_i;\varepsilon) = \underline{0} \label{dlm}
\end{equation}
and the impurity and excess scattering matrices are given, respectively, by
\begin{equation}
  \underline{D}_i(\hat{e}_i;\varepsilon) = \Big[\underline{1} - \Big(\underline{\tilde t}_i\inv{1}(\varepsilon) - \underline{t}_i\inv{1}(\hat{e}_i;\varepsilon)\Big)\,\underline{\tilde\tau}_{ii}(\varepsilon)\Big]\inv{1}
\end{equation}
and
\begin{equation}
  \underline{X}_i(\hat{e}_i;\varepsilon) = \Big[\Big(\underline{\tilde t}_i\inv{1}(\varepsilon) - \underline{t}_i\inv{1}(\hat{e}_i;\varepsilon)\Big)\inv{1} - \underline{\tilde\tau}_{ii}(\varepsilon)\Big]\inv{1}
\end{equation}

This leads to the desired expression for the Weiss fields, making use of the effective medium quantities:
\begin{equation}
  \vec{h}_i = \frac{1}{\pi}\,\ImTr\!\int\!\!\ud\varepsilon\,f(\varepsilon;\nu)\left[\frac{3}{4\pi}\!\int\!\!\ud\hat{e}_i\,\hat{e}_i\log\underline{D}_i(\hat{e}_i;\varepsilon)\right]
\end{equation}
where $f(\varepsilon;\nu)$ is the Fermi--Dirac distribution for energy $\varepsilon$ and chemical potential $\nu$. This expression has the form of an impurity integrated DOS, and in fact is derived from Lloyd's formula\cite{Staunton_TempAnis}, Eq. (\ref{lloyd}).

Now that we presented our construction of the DLM, we can proceed and perform a linear response analysis of how the DLM state responds to an inhomogeneous infinitesimal external magnetic field applied in each site. This procedure will generate our desired spin--spin correlation functions, which contain information about the coupling between magnetic moments, determined by the self--consistent effective medium.

\subsection{Linear response and magnetic susceptibility}
We begin by formally introducing an infinitesimal external magnetic field on each site, $\delta\vec{H}_i$. The Weiss field on site $i$, $\vec{h}_i$, will then be modified as $\vec{h}_i \rightarrow \vec{h}_i + \delta\vec{H}_i + \delta\vec{h}_i$. The small correction $\delta\vec{h}_i$ is due to the adjustment of the self--consistent effective medium in the presence of the infinitesimal external fields. In the absence of this feedback the magnetic response would be that of a system of non--interacting Langevin spins, as we will now show.

The response of the local magnetisation to the external field is by definition the magnetic susceptibility. In our linear response approach it is given by
\begin{equation}
  \frac{\delta\vec{m}_i}{\delta\vec{H}_j} \equiv \TENSOR{\chi}{ij}{} = \TENSOR{\chi}{0,i}{}\cdot\left(\delta_{ij}\TENSOR{\mathcal{I}}{}{} + \frac{\delta\vec{h}_i}{\delta\vec{H}_j}\right) = \TENSOR{\chi}{0,i}{}\cdot\TENSOR{\mathcal{W}}{ij}{}
\end{equation}
with the magnetic susceptibility of non--interacting moments defined as before. All probability distributions and effective medium quantities are those of the reference state, due to the linear response approach being used.

The coupling between the local moments is then contained in the response of the local Weiss fields to the infinitesimal external fields. Using the definition we obtain, after some algebra,
\begin{equation}
  \TENSOR{\mathcal{W}}{ij}{\alpha\beta} = \delta_{ij}\delta_{\alpha\beta}
- \Tr\frac{3}{4\pi}\!\int\!\!\ud\hat{e}_i\,\hat{e}_i^\alpha\,\underline{X}_i(\hat{e}_i)\sum_{k \neq i} \underline{\tilde \tau}_{ik}\frac{\delta\underline{\tilde t}_k\inv{1}}{\delta\vec{H}_j^\beta}\,\underline{\tilde \tau}_{ki}
\end{equation}
where the energy integration, taking the imaginary part and dividing by $\pi$ is being implied along with the trace.

A second equation can be obtained from the equation that defines the DLM effective medium, Eq. (\ref{dlm}), by obtaining the first--order correction to the effective medium $t$--matrix. Introducing the expansion
\begin{equation}
  \frac{\delta\underline{\tilde t}_i\inv{1}}{\delta\vec{H}_j} = -\beta\sum_k\vec{\underline{\lambda}}_{ik}\cdot\TENSOR{\mathcal{W}}{kj}{}
\end{equation}
which defines a `full' vertex, $\vec{\underline{\lambda}}_{ij}$, the following equation of motion can be written:
\begin{equation}
  \vec{\underline{\lambda}}_{ij} = \vec{\underline{\lambda}}_i^0\,\delta_{ij} + \!\int\!\!\ud\hat{e}_i\,P_i(\hat{e}_i)\,\underline{X}_i(\hat{e}_i)\sum_{k \neq i} \underline{\tilde \tau}_{ik}\,\vec{\underline{\lambda}}_{kj}\,\underline{\tilde \tau}_{ki}\,\underline{X}_i(\hat{e}_i) \label{fullv}
\end{equation}
by defining the `bare' vertex
\begin{equation}
  \vec{\underline{\lambda}}_i^0= \int\!\!\ud\hat{e}_i\,P_i(\hat{e}_i)\left(\hat{e}_i - \vec{m}_i\right)\underline{X}_i(\hat{e}_i) \label{barev}
\end{equation}

This gives for the magnetic susceptibility the expected expression
\begin{equation}
  \TENSOR{\chi}{ij}{} = \TENSOR{\chi}{0,i}{} + \frac{\beta}{3}\sum_k\TENSOR{\mathcal{S}}{ik}{}{}\cdot\TENSOR{\chi}{kj}{} \label{susc}
\end{equation}
where the spin--spin correlation function is given by
\begin{align}
  \TENSOR{\mathcal{S}}{ij}{\alpha\beta} = \frac{3}{\pi}&\,\ImTr\!\int\!\!\ud\varepsilon\,f(\varepsilon;\nu)\,\vec{\underline{A}}_i^\alpha(\varepsilon)\sum_{k \neq i} \underline{\tilde \tau}_{ik}(\varepsilon)\,\vec{\underline{\lambda}}_{kj}^\beta(\varepsilon)\,\underline{\tilde \tau}_{ki}(\varepsilon) \nonumber\\
\text{with}\,\; \vec{\underline{A}}_i(\varepsilon&) = \frac{3}{4\pi}\!\int\!\!\ud\hat{e}_i\,\hat{e}_i\,\underline{X}_i(\hat{e}_i;\varepsilon) \label{s2}
\end{align}
If relativistic effects are not included in the theory, then this expression becomes independent of the Cartesian indices, and thus isotropic in real space. This is a consequence of the previous discussion on the form of the single site $t$--matrices with and without SOC effects.

From the corresponding expressions, Eqs. (\ref{fullv}) and (\ref{barev}), it can be seen that the `bare' vertex is the direct response to the external applied fields, while the `full' vertex also incorporates fluctuation terms, defined by the angular averages in Eq. (\ref{fullv}). For the systems being studied in this paper these are small corrections, and so we can approximate the `full' vertex by its first iteration, by replacing it by the `bare' vertex on the right--hand--side of Eq. (\ref{fullv}).

Focusing on the high temperature PM state, and make use of its high symmetry to take the lattice Fourier transform of Eq. (\ref{susc}), and inverting to solve for the susceptibility:
\begin{equation}
  \TENSOR{\chi}{}{}(\vec{q}) = \mu^2 \Big[\,3 k_{\text{B}}T\,\TENSOR{\mathcal{I}}{}{} - \TENSOR{\mathcal{S}}{}{}(\vec{q})\,\Big]\inv{1}
\end{equation}
where the magnitude of the local moments in the PM state, $\mu$, was reintroduced. This shows that the spin--spin correlation function plays a similar role to that of the Fourier transform of the exchange constants in the familiar Heisenberg model. The analysis that was carried out for that case can then be mapped to the present description, and be used to interpret our results. It should be noted, however, that the spin--spin correlation function is derived from the electronic structure in the DLM state, and incorporates fluctuation effects.

The core of Eq. (\ref{s2}) is the convolution
\begin{equation}
  \sum_{k} \underline{\tilde \tau}_{ik}\,\vec{\underline{\lambda}}_{kj}\,\underline{\tilde \tau}_{ki} \longrightarrow \!\int\!\!\frac{\ud\vec{k}}{\Omega_{\text{BZ}}}\,\underline{\tilde \tau}(\vec{k}\!+\!\vec{q})\,\vec{\underline{\lambda}}(\vec{q})\,\underline{\tilde \tau}(\vec{k})
\end{equation}
which shows that Fermi surface nesting effects are accounted for, through the product of the scattering path operators. Thus RKKY behaviour may occur\cite{Hughes_Nature}.

We will now present the results of our calculations for Mn$_1$/X(111), X = Pd, Pt, Ag and Au. This choice of closely related substrates will highlight the role of hybridisation with the substrate in determining the predicted magnetic instability, as well as the role of varying SOC on the anisotropic effects.

\section{Computational details}
Our calculations are carried out in the framework of the Screened Korringa-Kohn-Rostoker\cite{Szunyogh_SKKR} (SKKR) method for layered systems. The relativistic formalism was adopted, in the Local Spin Density Approximation (LSDA), using the atomic sphere approximation. The energy integrations were performed on an asymmetric semicircular contour in the upper half of the complex plane using 24 points.

The self--consistent paramagnetic DLM potentials were generated in the scalar--relativistic approximation, using 200 k--points in the irreducible wedge of the 2D Brillouin zone, which has $C_{3v}$ symmetry. These potentials were then used in the fully relativistic linear response calculations, with 3721 k--points in the full 2D Brillouin zone ensuring all results for the spin--spin correlation function are converged within 0.1 meV. This is possible as only complex energies with finite imaginary part are used in the energy integration.

The geometry used in our calculations was as follows. The interface region was composed of eight layers of the transition metal substrate, the Mn ML, and three layers of empty spheres. This was then matched to the semi--infinite bulk substrate and to the semi--infinite vacuum region, so that no slab approximation was used.

No attempt was made at determining the equilibrium geometries for the four systems studied. However, the influence of different interlayer spacings between the magnetic monolayer and the substrate was investigated for the cases of Mn/Ag and Mn/Au, and it was found that there was no qualitative change in the magnetic properties, up to 15\% inward relaxation. The dependence of the magnetic properties on the in--plane lattice constant was addressed in a previous study\cite{Hobbs_MnCrCu111}, and for the range relevant to our systems the same conclusion can be made. On the other hand, the relaxed geometrical parameters depend on the magnetic state and other approximations, such as the exchange--correlation functional and the handling of the charge density, so a definite statement is precluded. We thus progress with a simple model of the geometric structure, which we now describe.

To aid comparison, the average of the experimental lattice constants for Pd and Pt was used for both Pd and Pt (a = 3.905 \AA), and likewise for Ag and Au (a = 4.085 \AA). This means that the in--plane lattice constant for the Mn ML is the same for the Pd and Pt substrates, and also for the Ag and Au substrates. To estimate the inward relaxation, it was assumed that Mn grows on Cu(111) without any significant relaxation, thus defining the Mn ML hard sphere radius. By considering how these hard spheres stack on top of the larger ones corresponding to the substrate, the following relation was obtained for the ratio between the estimated and the ideal distance between planes: $d/d_0 = \sqrt{3/8\,(1 + a_{\text{Cu}}/a_{\text{X}})^2 - 1/2}$. We then round the results and use 5\% inward relaxation for the Pd and Pt substrates, and 10\% for Ag and Au.

\begin{figure}[t]
\centering
\includegraphics[width=\columnwidth]{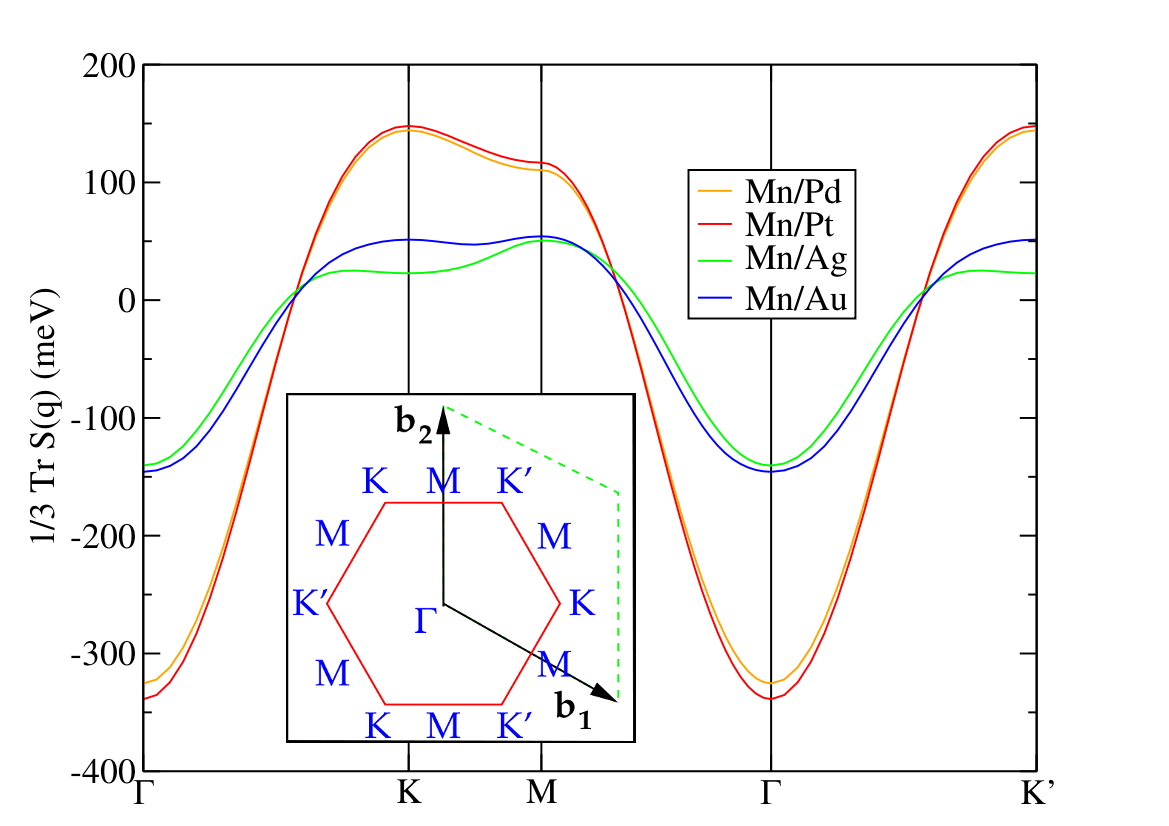}
\caption{\label{isoq} The isotropic part of the $\TENSOR{\mathcal{S}}{}{}(\vec{q})$ tensor along high symmetry lines of the 2D BZ. The maxima signal the favoured magnetic instability, which correspond to the row--wise AF (Ag, Au) or triangular N\'eel states (Pd, Pt), see also Fig. \ref{magq}.}
\end{figure}

\section{Anisotropic spin--spin correlations in M\lowercase{n} monolayers}
We begin by presenting our results for the anisotropic spin--spin correlations in Mn$_1$/X(111), with X = Pd, Pt, Ag and Au, and then proceed to a discussion and comparison with other theoretical and experimental works. The same decomposition that was used in Eq. (\ref{decomp}) will be used to extract the information from the spin--spin correlation function.

\subsection{Instabilities of the paramagnetic state}

\begin{figure}[t]
\includegraphics[width=0.75\columnwidth]{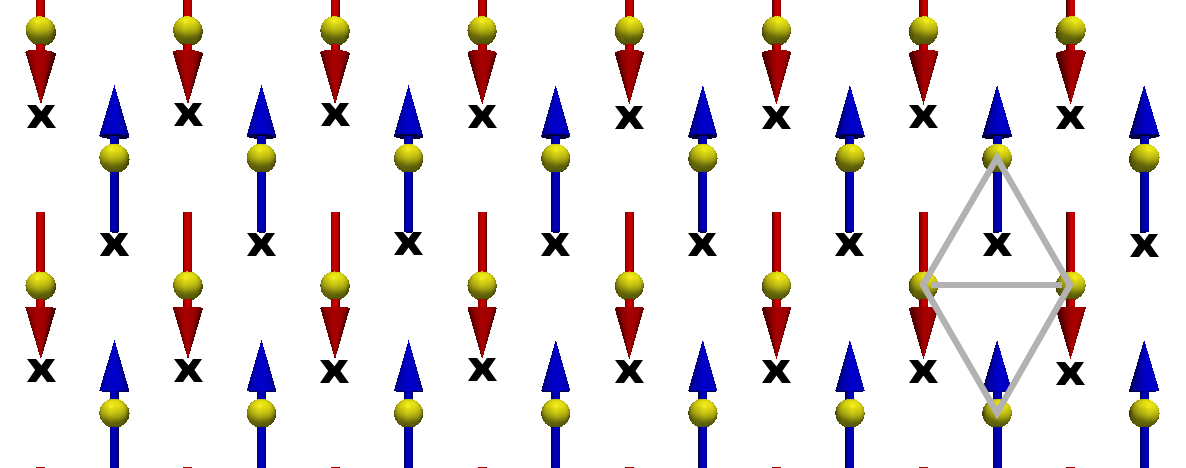}
\rule{0.85\columnwidth}{0.1em}\vspace{1em}
\includegraphics[width=0.75\columnwidth]{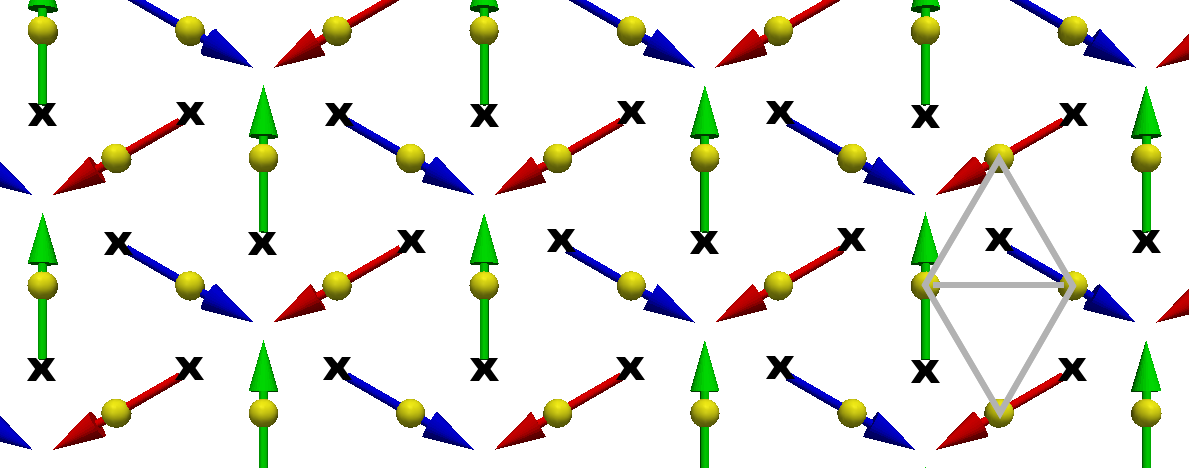}
\rule{0.85\columnwidth}{0.1em}\vspace{1em}
\includegraphics[width=0.75\columnwidth]{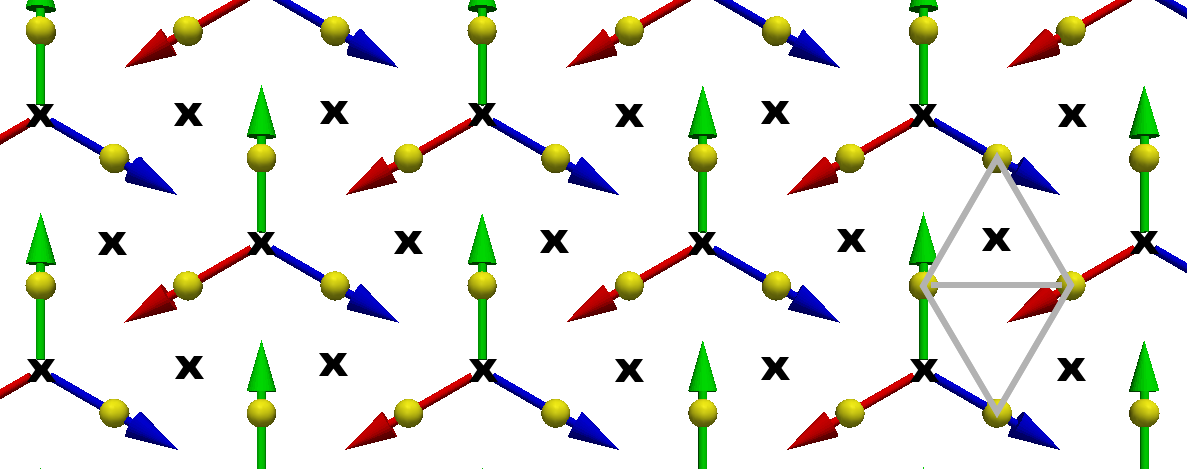}
\caption{\label{magq} Magnetic structures associated with high--symmetry points on the 2D BZ. Top: row--wise AF state associated with the M point. Middle and bottom: triangular N\'eel states of opposite chiralities --- see discussion in text. Spheres show the positions of the magnetic atoms. Crosses mark the positions of the nearest substrate atoms. `Up' and `down' triangles are shown, with respect to the nearest substrate atoms.}
\end{figure}
\begin{table}[b]
\caption{Calculated real--space isotropic interactions, $\TENSOR{\mathcal{S}}{ij}{\text{I}}$, and magnitude of DM vectors, $|\vec{D}_{ij}|$, for first and second NNs.\label{summary}}
\begin{ruledtabular}
\begin{tabular}{c c c c c c}
& & Mn/Pd & Mn/Pt & Mn/Ag & Mn/Au \\
\hline
\multirow{2}{*}{$\TENSOR{\mathcal{S}}{ij}{\text{I}}$} & First shell & -53.33 & -55.49 & -18.51 & -22.92 \\
 & Second shell & -2.50 & -2.72 & -5.91 & -3.22 \\
 \hline
 \multirow{2}{*}{$|\vec{D}_{ij}|$} & First shell & 1.19 & 4.41 & 0.29 & 0.34 \\
  & Second shell & 0.34 & 0.27 & 0.00 & 0.28
\end{tabular}
\end{ruledtabular}
\end{table}

First we discuss the isotropic part of the tensor, $\TENSOR{\mathcal{S}}{}{\text{I}}(\vec{q})$. This can be conveniently plotted along high--symmetry lines in the 2D Brillouin zone, and can be used as a guide to the favoured magnetic instability, given by the modulation vector associated with its maximum value, see discussion following Eq. (\ref{free}). For the Pd and Pt substrates, the favoured magnetic instability is associated with the K and K' points, which corresponds to triangular N\'eel states, while for the Ag substrate the favoured instability is associated with the M point, which corresponds to the row--wise AF state. The case with the Au substrate appears to show an approximate degeneracy between the K and M point--type of AF.

The high symmetry magnetic states are illustrated in Fig. \ref{magq}. Considering only the information in $\TENSOR{\mathcal{S}}{}{\text{I}}(\vec{q})$ , as K is mapped into K' by a mirror symmetry, they are equivalent, and degenerate in energy. Nothing can be said about the real space orientation of the local moments, only about the relative angles between them, as the tensor is isotropic in real space.

Looking at the real--space values of the tensor, see Table \ref{summary} (a negative sign means an AF coupling), it can be seen that the contribution from the nearest--neighbours (NNs) is dominant for Pd and Pt and, though smaller in magnitude, also for Ag and Au. These interactions leads to the competition between magnetic states shown in Fig. \ref{isoq}, for the Au substrate. This competition effect is similar to the behaviour of the $J_1/J_2$ isotropic Heisenberg model in a triangular lattice\cite{Heinze_MnAg}.

\subsection{Real space magnetic structure}

The anisotropic part of $\TENSOR{\mathcal{S}}{}{}(\vec{q})$ is small compared to the isotropic part, but it sets the real space orientation of the spins. The largest eigenvalue of $\TENSOR{\mathcal{S}}{}{}(\vec{q})$ has local extrema at the same positions as $\TENSOR{\mathcal{S}}{}{\text{I}}(\vec{q})$, see Fig \ref{isoq}, so the magnetic instabilities remain unchanged. For the systems studied, the symmetric anisotropy $\TENSOR{\mathcal{S}}{}{\text{S}}(\vec{q})$ favours an in--plane spin configuration.

\begin{figure}[t]
\centering
\includegraphics[height=0.75\columnwidth]{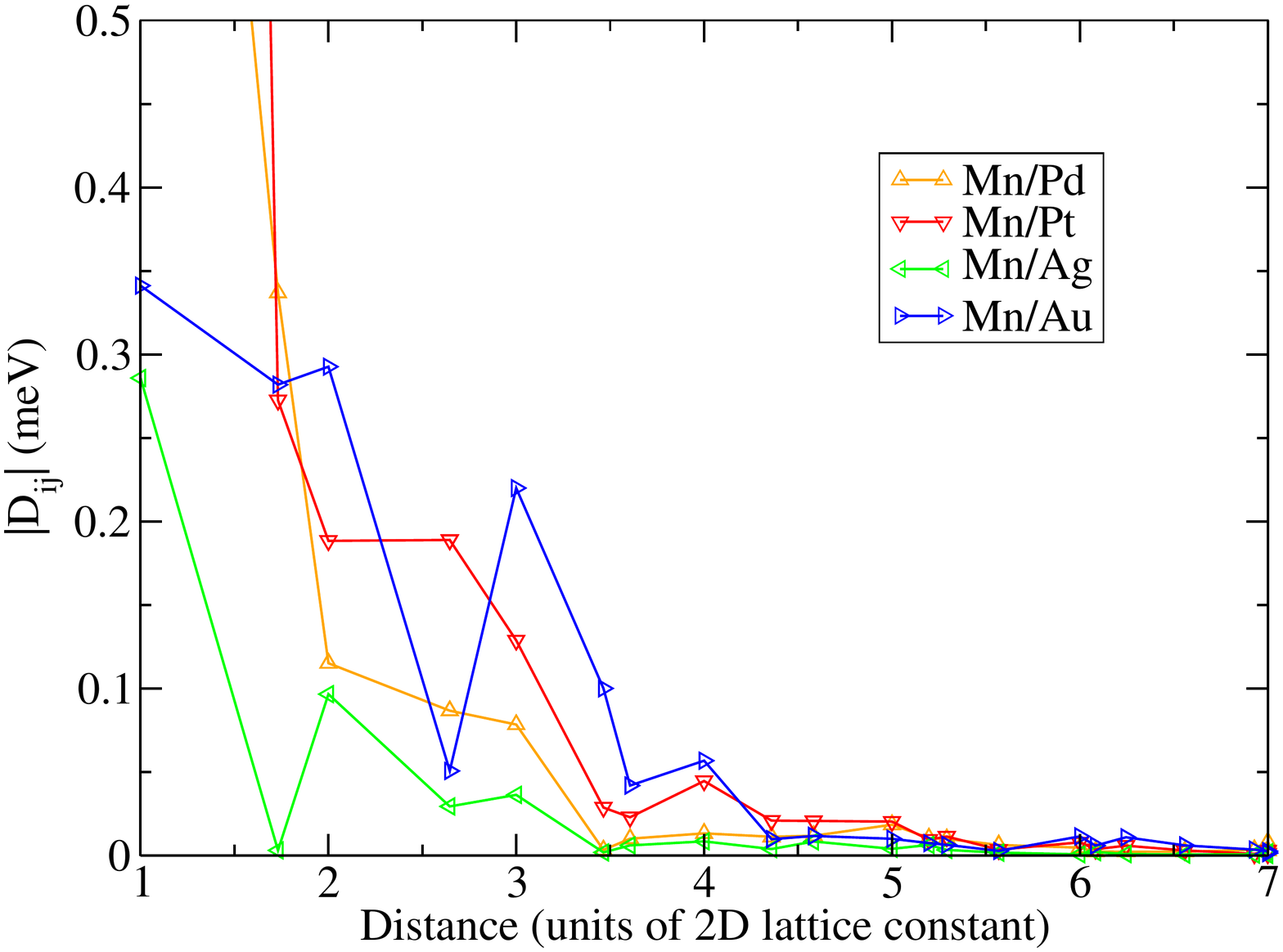}
\caption{\label{dmlen} The absolute value of the DM vector against distance. In the Pd and Pt substrates the DM vector for the first shell of neighbouring atoms is greatly enhanced, see Table \ref{summary}. The existence of sizable DM vectors over several shells is remarkable.}
\end{figure}

\begin{figure}[t]
\centering
\includegraphics[height=0.75\columnwidth]{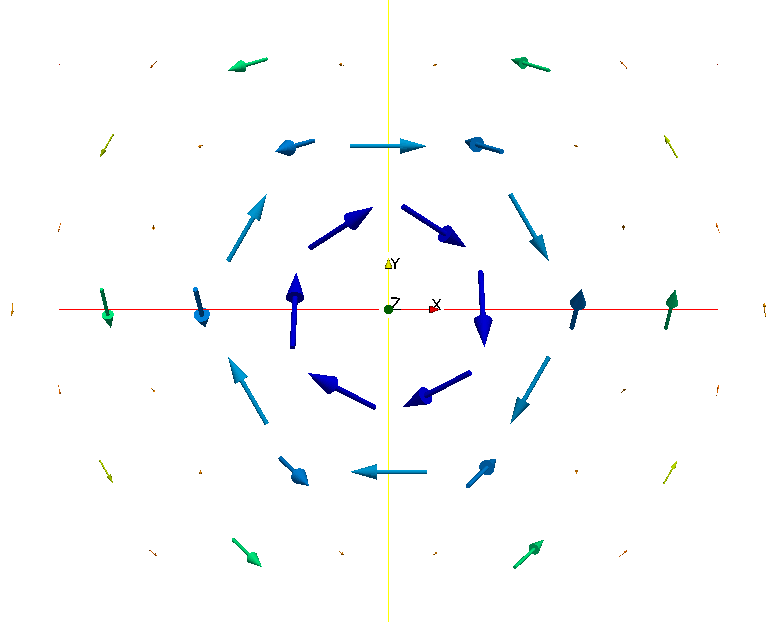}
\caption{\label{dmshell} The DM anisotropy field for Mn/Au. Vectors of the same colour have the same magnitude, and belong to the same lattice shell, representing the interaction with the central site. Note that $z$--component is small for most DM vectors, and so is not very apparent in this 3D plot.}
\end{figure}

The antisymmetric part of the tensor, $\TENSOR{\mathcal{S}}{}{\text{A}}(\vec{q})$, can be cast in the form of a DM vector, see Eq. (\ref{dmdef}). The size of this DM vector is the same for all atoms in the same lattice shell, and is plotted as a function of the distance between the atoms in Fig. \ref{dmlen}. As expected, it is largest for the first shell, and also greatly enhanced in the Pd and Pt substrates. The case of the Au substrate is most interesting, as there are several lattice shells with comparable DM vectors (see Fig. \ref{dmshell}), leading to a complex unidirectional anisotropy field. In the $\TENSOR{\mathcal{S}}{}{}(\vec{q})$ picture, this anisotropy leads to a splitting of the eigenvalues, as exemplified in Eq. (\ref{example}).

If the triangular N\'eel state is the favoured magnetic instability, the presence of the DM anisotropy will select one of the two possible chiralities. This is the case for the Pd and Pt substrates. We define a chirality vector in the following way\cite{Szunyogh_IrMn3}:
\begin{equation}
  \vec{\xi} = \delta\vec{m}_1 \times \delta\vec{m}_2 + \delta\vec{m}_2 \times \delta\vec{m}_3 + \delta\vec{m}_3 \times \delta\vec{m}_1
\end{equation}
Here the numbers label the magnetic atoms going anticlockwise around each triangle of NNs in the lattice. There are two types of triangles in the magnetic layer: those that do encircle an atom of the nearest substrate layer, which will be called `up', and those which do not, called `down'. These two types of triangles are marked in Fig. \ref{magq}. For collinear magnetic states the chirality vector is obviously zero (see top panel of Fig. \ref{magq}).

Referring to the middle panel of Fig. \ref{magq}, for all `up' triangles $\vec{\xi} \propto -\hat{z}$ (the `spins' rotate in a clockwise fashion), and all `down' ones $\vec{\xi} \propto \hat{z}$ (the `spins' rotate in an anticlockwise fashion). This pattern is reversed for the magnetic structure depicted in the bottom panel. The net chirality is thus zero for both magnetic states. However, these two chirality patterns are clearly distinguishable, as the `up' and `down' triangles are inequivalent. This is encoded in the $C_{3v}$ symmetry of the system, and allows the lifting of the chiral degeneracy.

It can be shown straightforwardly that the high symmetry of the triangular N\'eel state dictates that the energy difference between the two chiral states is set by the $z$--component of the DM interaction\cite{Udvardi_DMcoupling}. For the systems studied, the DM vectors are mostly in--plane, thus their $z$--component is small, as can be seen for example in Fig. \ref{dmshell}. This means the lifting of the chiral degeneracy is smaller than would be anticipated, from considerations based on the length of the DM vectors. The energy difference between the two magnetic pattenrs depicted in the lower panels of Fig. \ref{magq} is of the order of 1 meV in all cases studied. It is not large enough to favour the triangular N\'eel state, when the isotropic part of the tensor favours the row--wise AF state, as in the Mn/Ag and Mn/Au cases. If the dominant isotropic interactions were to favour another noncollinear state with less symmetry, as was demonstrated for Mn$_1$/W(001), then this splitting would be more significant\cite{Ferriani_MnW001}.

\subsection{Connecting magnetism and electronic structure}

We now proceed to establish a link between the electronic structure and the just described magnetic properties.

\begin{figure}[t]
\centering
\includegraphics[width=\columnwidth]{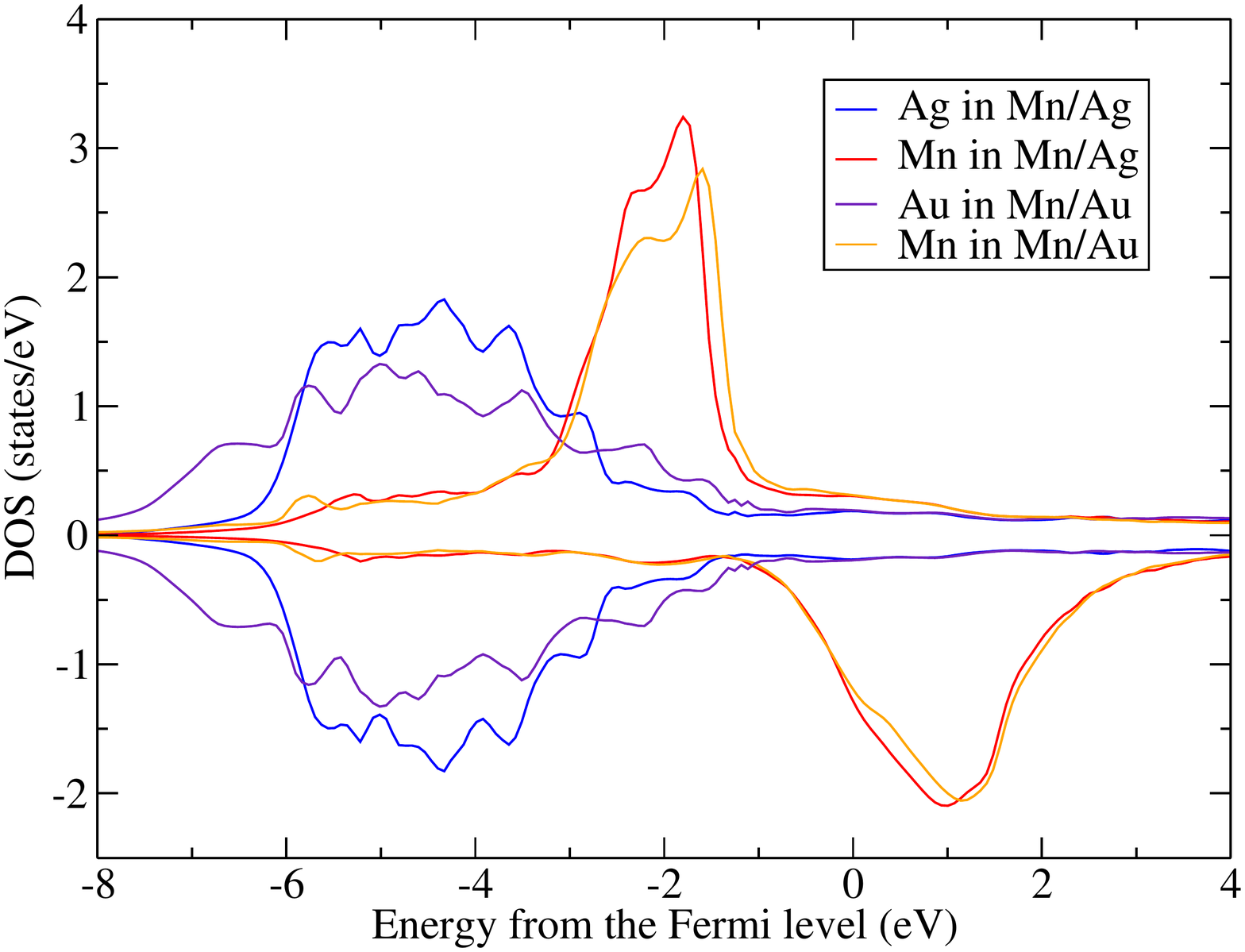}
\includegraphics[width=\columnwidth]{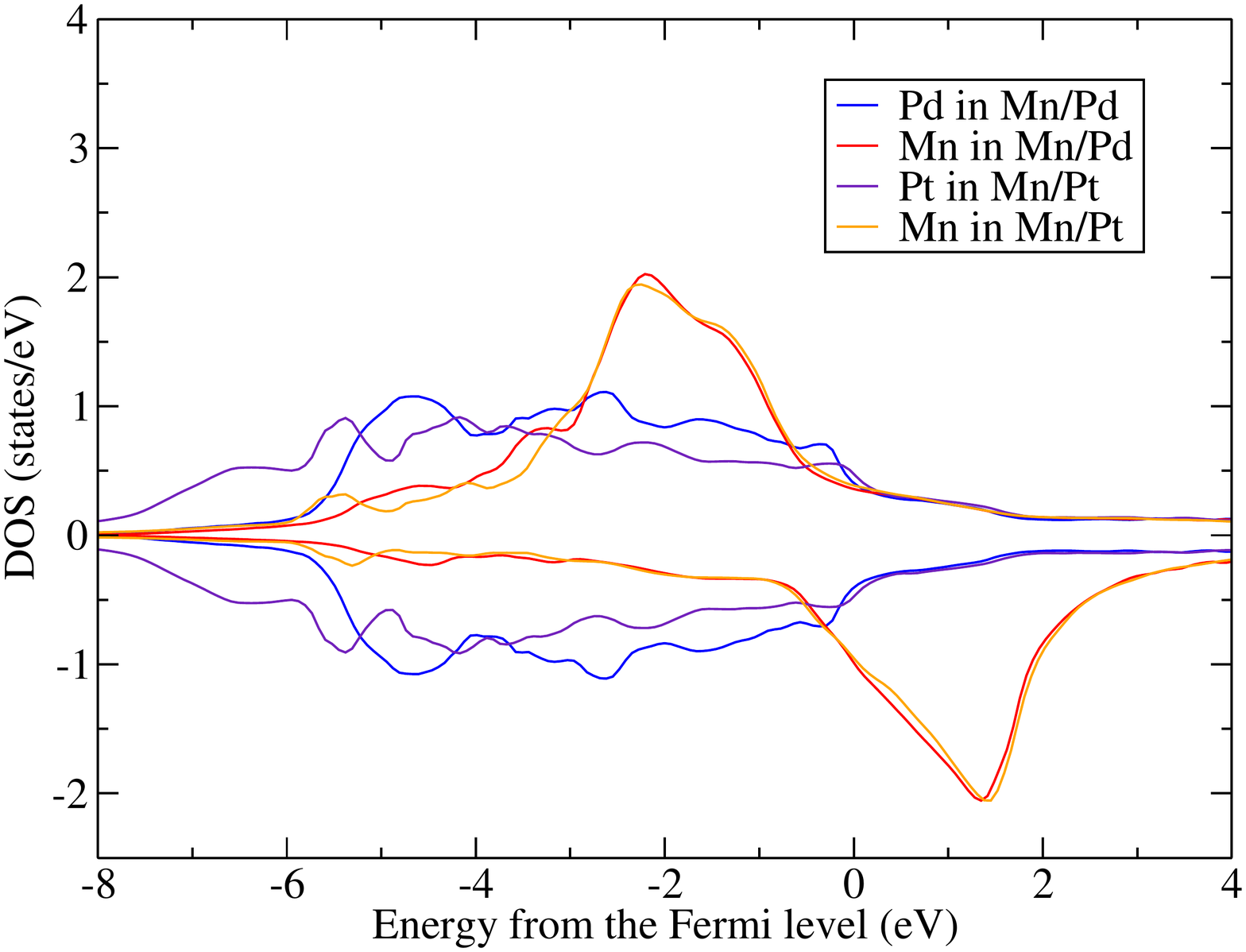}
\caption{\label{ldos} The layer--resolved DOS for Mn$1$/X(111), with X = Pd, Pt, Ag and Au. Only the DOS for the Mn and the nearest substrate layer is shown.}
\end{figure}

The influence of the substrate on the electronic structure of the Mn ML can be readily understood from the layer--resolved DOS, in the DLM state, shown in Fig. \ref{ldos}. The Mn minority states are just slightly affected by changing the substrate, as can be seen be comparing all four cases. The majority states, however, show a marked difference. For the Ag and Au substrates these states are fairly narrow, while for the Pd and Pt substrates they are much wider in energy, and show strong signs of hybridisation. This can be explained by the position of the substrate d--states. Ag and Au have filled d--bands, lower in energy than the Mn majority states, and so there is little hybridisation. On the other hand, Pd and Pt have partially filled d--bands, which extend up to the Fermi energy, so there is much stronger hybridisation with the Mn majority states. The strength of the hybridisation with the substrate alo affects the size of the DLM spin moments: for Mn/Ag and Mn/Au it is about 3.9 $\mu_\text{B}$, in agreement with previous calculations, while for Mn/Pd and Mn/Pt it drops to about 3.6 $\mu_\text{B}$.

A simple picture can be invoked to synthesise these arguments: on the Ag and Au substrates, the magnetic properties of the Mn ML are quite similar to those of a conceptual free standing ML, as found before in other electronic structure calculations. This explains the similarity between the $\TENSOR{\mathcal{S}}{ij}{\text{I}}$ tensors for the Ag and Au substrates, see Table \ref{summary}. The Pd and Pt substrates then represent a strong enhancement of the NN interactions due to hybridisation with the substrate.

Referring also to Fig. \ref{dmlen}, it can be seen that the anisotropic effects scale with the atomic number of the substrate, due to the SOC effect, as expected. The DM vectors on the first shell, however, show a marked difference: they are much larger in the Pd and Pt substrates than in the Ag and Au substrates, regardless of the similarity between atomic numbers. This reveals another ingredient to a large anisotropic effect: strong hybridisation with the substrate. This is why the first shell DM vector in Mn/Pd is three times larger than the one in Mn/Au, despite the differences in atomic number, and so in SOC strength.

\subsection{Mn$_1$/Ag(111): comparison and discussion of previous theoretical and experimental work}
Now that we presented our results for the four systems studied, we wish to focus on a more detailed comparison for the most studied case, that of Mn$_1$/Ag(111).

It was found experimentally that the magnetic ground state was the triangular N\'eel state\cite{Gao_MnAg}, in contrast with the row--wise AF state predicted by our theory, and also by previous electronic structure calculations\cite{Kruger_MnAg,Heinze_MnAg}. We suggest two explanations for the disagreement between theory and experiment. Both are based on the delicate balance found between the two types of AF:
\begin{list}{\labelitemi}{\leftmargin=1em}
  \item There may be some buried Mn atoms underneath the studied Mn triangular islands. This would enhance the hybridisation to the substrate, which has a very strong influence on the strength of the NN couplings, as found for the Pd and Pt substrates, and thus favours the triangular AF state.
  \item There may be an inadequate treatment of the exchange--correlation effects in the Mn ML. Although this might be a small correction, the Mn$_1$/Ag(111) system, as described by the LSDA, has a small energy difference between the two types of AF states just discussed. The electronic structure changes that may be brought about by an improved treatment might be enough to tilt the preference from the row--wise to the triangular AF state.
\end{list}

The experimental data on Mn$_1$/Ag(111) also suggests that similarly oriented magnetic islands with respect to the substrate have different magnetic domains, and so different magnetic anisotropies. We propose that this might be due to two different chirality patterns of the triangular AF state. As explained by the experimentalists, similarly oriented islands may grow on the substrate in an fcc or hcp--like stacking. Going back to our discussion on chirality, the difference between the two cases is whether the nearest substrate atoms are in the centre of the `up' or `down' triangles. Assuming that the `spins' around the substrate atoms always have the same chirality vector, the two different ways in which the Mn atoms can stack on the substrate lead naturally to two distinct chirality patterns, and so two different magnetic domains.

\section{Conclusions}
We presented a new theory of the anisotropic spin--spin correlations in good local moment systems, and demonstrated its application to Mn$_1$/X(111), with X = Pd, Pt, Ag and Au. The kind of magnetic structure for which the PM state of each system is unstable was identified. Mn/Pd and Mn/Pt should order in the triangular N\'eel state, while Mn/Ag and Mn/Au order in the row--wise AF state. For the last two systems it was found that the energy difference between the two types of AF is quite small. The symmetric part of the anisotropy dictates that these should be in--plane magnetic configurations, while the antisymmetric, or DM--type, anisotropy lifts the chiral degeneracy of the triangular N\'eel state.

For Mn$_1$/Ag(111), which is the most studied system in the literature, we found agreement with the findings of previous theoretical work, while experimental results point to a different magnetic state, namely the triangular N\'eel state. We suggested two resolutions for this inconsistency. Moreover, we proposed that the two types of magnetic domains found experimentally are manifestations of the chirality of the triangular AF state.

We also wish to highlight our observation that the anisotropic effects induced by the substrate do not simply scale with its atomic number. The strength of the hybridisation with the substrate may play a dominant role, and so a careful choice of substrate might yield similar anisotropic properties using lighter elements, thus dispensing the use of their heavier counterparts, which may be rarer and more expensive.

The theory presented here is more general than the applications chosen to illustrate it. In the interest of simplicity the formalism was presented for a single magnetic lattice or layer, but can be straightforwardly extended to multisublattice or multilayered systems. Our investigations are now progressing in this direction, with the Co/IrMn exchange--bias system, a spin valve element, as its driving application.

\begin{acknowledgments}
This work was supported by the  PhD grant SFRH/BD/35738/2007 awarded by FCT Portugal, using funding from FSE/POPH. This work was also supported by the Hungarian Research Foundation (contract OTKA K77771) and by the New Hungary Development Plan (Project ID: T\'AMOP-4.2.1/B-09/1/KMR-2010-0002). A collaboration visit was kindly sponsored by the Psi--k Network. We acknowledge the computational resources provided by the Centre for Scientific Computing at the University of Warwick, UK.
\end{acknowledgments}

\bibliography{bibdatabase}
\end{document}